\documentclass[conference]{IEEEtran}
\IEEEoverridecommandlockouts
\usepackage{cite}
\usepackage{amsmath,amssymb,amsfonts}
\usepackage{algorithmic}
\usepackage{booktabs}
\usepackage{graphicx}
\usepackage{textcomp}
\usepackage[table]{xcolor}
\usepackage{makecell}
\def\BibTeX{{\rm B\kern-.05em{\sc i\kern-.025em b}\kern-.08em
    T\kern-.1667em\lower.7ex\hbox{E}\kern-.125emX}}
\usepackage{eso-pic}
\usepackage[hidelinks]{hyperref}
\AddToShipoutPictureBG*{%
  \AtPageLowerLeft{%
    \put(\LenToUnit{0.75in},\LenToUnit{0.28in}){%
      \parbox{\dimexpr\paperwidth-1.5in\relax}{\scriptsize\centering
      \copyright\ 2026 IEEE. Personal use of this material is permitted. Permission from IEEE must be obtained for all other uses, in any current or future media, including reprinting/republishing this material for advertising or promotional purposes, creating new collective works, for resale or redistribution to servers or lists, or reuse of any copyrighted component of this work in other works. Accepted version of the paper published in the 2026 IEEE/ACM Conference on Connected Health: Applications, Systems and Engineering Technologies (CHASE), pp. 428--433, DOI: 10.1109/CHASE69719.2026.00073.}}}}
\begin{document}

\title{Reliability Stress Tests and Decision-Time Routing for Chest X-ray Vision-Language Models}

\author{\IEEEauthorblockN{Xinye Yang}
\IEEEauthorblockA{\textit{Department of Computer Science} \\
\textit{Brown University}\\
Providence, RI, USA \\
xinye\_yang@brown.edu}
\and
\IEEEauthorblockN{Zhusi Zhong}
\IEEEauthorblockA{\textit{Brown University} \\
\textit{Brown University Health}\\
Providence, RI, USA \\
zhusi\_zhong@brown.edu}
\and
\IEEEauthorblockN{Scott Collins}
\IEEEauthorblockA{\textit{Brown University} \\
\textit{Brown University Health}\\
Providence, RI, USA \\
scollins1@brownhealth.org}
\and
\IEEEauthorblockN{Grayson Baird}
\IEEEauthorblockA{\textit{Brown University} \\
\textit{Brown University Health}\\
Providence, RI, USA \\
grayson\_baird@brown.edu}
\and
\IEEEauthorblockN{Xuyu Wang}
\IEEEauthorblockA{\textit{School of Computing and Information Sciences} \\
\textit{Florida International University}\\
Miami, FL, USA \\
xuywang@fiu.edu}
\and
\IEEEauthorblockN{Zhicheng Jiao*}
\IEEEauthorblockA{\textit{Brown University} \\
\textit{Brown University Health}\\
Providence, RI, USA \\
zhicheng\_jiao@brown.edu \\
\textit{*Corresponding author}}
}

\maketitle

\begin{abstract}
Medical vision-language model (VLM) evaluation is sensitive to workflow design, prompting strategy, and benchmark construction, yet most studies treat these factors in isolation. We introduce a reliability stress test for chest X-ray interpretation built on two balanced datasets (a private report-backed set and a curated MIMIC subset). Three medical VLMs---CheXagent, MedGemma-4B, and MedGemma-27B---are evaluated across three prompt styles and two workflows (single-VLM and multi-agent), producing 36 configurations. We show that exact-match accuracy alone can overstate the effectiveness of conservative models that default to ``Normal'' predictions. Diagnostic reliability also depends heavily on model family and scale: multi-agent reasoning helps some configurations but hurts others. Building on these observations, we propose a decision-time routing framework that selectively escalates to multi-agent inference only when beneficial, improving the cost--quality trade-off over fixed workflows. Our results highlight the need for evaluation protocols that jointly consider prompt sensitivity, failure-mode diversity, and workflow choice before clinical deployment.
\end{abstract}


\section{Introduction}
Medical vision-language models (VLMs) have rapidly advanced automated chest X-ray interpretation, with the potential to reduce radiologist workload and widen access to expert-quality diagnostics. Yet as these systems move from benchmarks toward clinical prototypes, growing evidence shows that medical multimodal architectures are brittle under distribution shifts, complex reasoning tasks, and varied evaluation setups \cite{johnson2019mimic,royer2024multimedeval,sepehri2025mediconfusion}. Two evaluation pitfalls are frequently overlooked. First, changes in prompt design and workflow structure alter not just headline accuracy but the entire distribution of clinical errors a model produces. Second, adding advanced reasoning pipelines---multi-agent orchestration or chain-of-thought protocols---can improve diagnostic output for some cases while suppressing genuine pathological findings in others.

We investigate these challenges through a controlled reliability stress test and treat the observed workflow instability as a practical routing problem rather than a mere evaluation artifact. Instead of proposing a new foundation model, we focus on two questions that are central to safe clinical integration: \emph{(i) How do prompt style and multi-stage workflow design interact with model family and scale for chest X-ray interpretation? (ii) When should a system escalate from fast, low-cost single-VLM inference to a more deliberate multi-agent workflow?}

We evaluate CheXagent \cite{chen2024chexagent} alongside two MedGemma scales \cite{medgemma2025} on two balanced 50-case datasets. The primary dataset is a private cohort with paired clinical reports; the second is a curated MIMIC-CXR subset \cite{johnson2019mimic} that also provides matched report supervision. We test three prompt styles under two workflows: direct single-VLM inference and a lightweight, structured multi-agent pipeline.

Our main contribution is a controlled, multi-dimensional evaluation together with a deployable routing strategy. The experiments support four claims:
\begin{itemize}
\item Exact-match evaluation alone is misleading: conservative models hide weaknesses by over-predicting \texttt{Normal}.
\item System reliability depends strongly on model family and scale, not just prompt wording.
\item Multi-agent reasoning provides conditional, not universal, performance gains.
\item Decision-time routing is a practical deployment strategy: selective escalation consistently improves cost--quality trade-offs over fixed workflows.
\end{itemize}

\section{Related Work}
Our work draws on three lines of research. First, domain-specific medical VLMs---LLaVA-Med, CheXagent, MedGemma \cite{li2023llavamed,chen2024chexagent,medgemma2025}---are typically evaluated under fixed, single-pass inference, leaving open the question of how workflow variations interact with model scale in practice.

Second, evaluation frameworks such as MultiMedEval \cite{royer2024multimedeval} and XBench \cite{luo2025xbench} cover broad visual-language grounding but often miss the interactive, multi-step nature of real-world diagnosis. We trade breadth for depth, stress-testing prompt robustness and multi-stage workflow reliability on a focused clinical task.

Third, reliability benchmarks like MediConfusion \cite{sepehri2025mediconfusion} and PSF-Med \cite{sadanandan2026psfmed} have revealed model brittleness through paraphrase sensitivity and hallucination rates. We extend these findings from static prompt perturbation to dynamic workflow behavior, showing that performance can shift substantially when the same model weights encounter different instructions or multi-agent pipelines.

Separately, inference routing addresses cost-efficient deployment of large models. Building on FrugalGPT \cite{chen2023frugalgpt}, methods like AT-CXR \cite{li2025atcxr} explore uncertainty-aware triage, while benchmarks such as VL-RouterBench \cite{huang2025vlrouterbench} and routing frameworks \cite{vassef2025onevlmtworoles,zhang2026baar,liu2026caster,lou2025adacot} formalize cost--quality trade-offs. Unlike prior work that routes between distinct models or human experts, we frame routing as a binary decision-time choice---fast single-VLM inference versus deliberate multi-agent processing---using frozen model parameters.

\section{Experimental Setup}
\subsection{Datasets}
We use two evaluation cohorts, each with 50 chest X-ray studies evenly split between normal and abnormal cases. The private dataset's labels come directly from the original clinical reports. The public MIMIC subset was curated from MIMIC-CXR \cite{johnson2019mimic} to ensure that every included case has a matched report. Table~\ref{tab:data_summary} summarizes the dataset configurations.

\begin{table}[t]
\caption{Dataset Summary.}
\label{tab:data_summary}
\centering
\small
\begin{tabular}{lcccc}
\toprule
\textbf{Dataset} & \textbf{Cases} & \textbf{Normal} & \textbf{Abnormal} & \textbf{Reports} \\
\midrule
Private formal & 50 & 25 & 25 & 50 \\
MIMIC formal & 50 & 25 & 25 & 50 \\
\bottomrule
\end{tabular}
\end{table}

The MIMIC subset provides a reproducible, public baseline with matched reports, reducing the risk that label-source mismatches distort the workflow comparisons.

\begin{figure*}[t]
\centering
\includegraphics[width=0.8\textwidth]{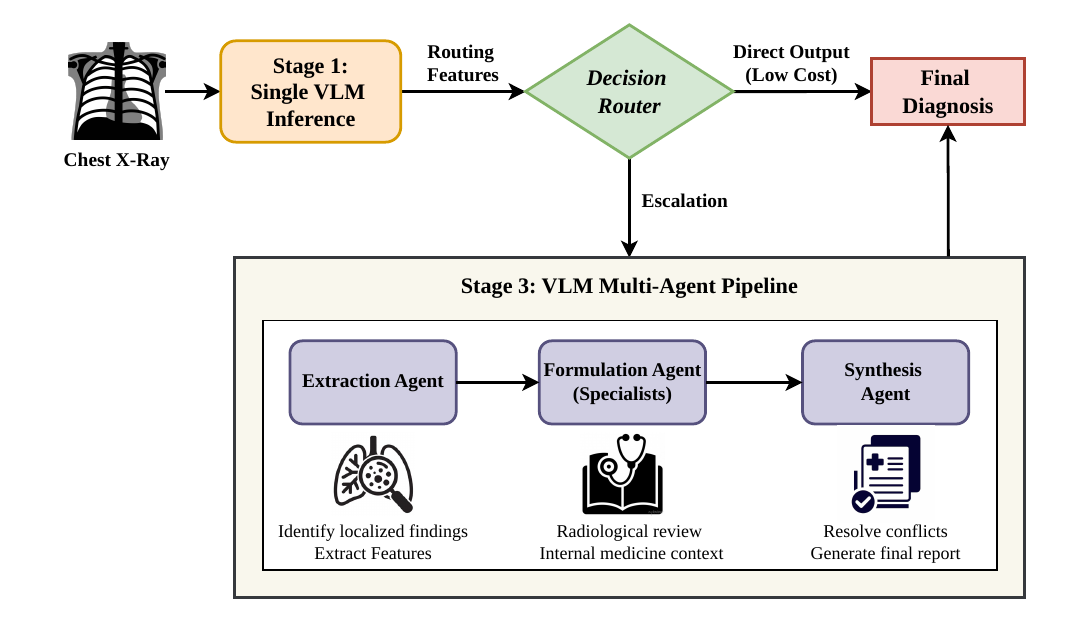}
\caption{Decision-time routing framework for chest X-ray interpretation. A single-VLM pass produces initial predictions and routing features. The decision router then either returns the low-cost output directly or escalates the case to a multi-agent pipeline with extraction, formulation, and synthesis stages.}
\label{fig:workflow}
\end{figure*}

\begin{table*}[t]
\caption{Comprehensive Formal Results on the Curated MIMIC Subset.}
\label{tab:mimic_results}
\centering
\scriptsize
\setlength{\tabcolsep}{4.0pt}
\renewcommand{\arraystretch}{1.08}
\rowcolors{3}{gray!8}{white}
\resizebox{\textwidth}{!}{
\begin{tabular}{lllrrrrrrrr}
\toprule
\textbf{Model} & \textbf{Prompt} & \textbf{Workflow} & \makecell{\textbf{Exact}} & \makecell{\textbf{Partial}} & \textbf{Macro F1} & \makecell{\textbf{Hall FP}} & \makecell{\textbf{Normal Acc}} & \makecell{\textbf{Abn Partial}} & \makecell{\textbf{Latency (s)}} & \textbf{Tokens} \\
\midrule
CheXagent & structured & single & 6.0\% & 30.0\% & 0.118 & 0.351 & 12.0\% & 48.0\% & 1.32 & 22.2 \\
CheXagent & structured & multi & 24.0\% & 32.0\% & 0.160 & 0.089 & 40.0\% & 24.0\% & 6.30 & 144.5 \\
CheXagent & concise & single & 6.0\% & 28.0\% & 0.093 & 0.380 & 12.0\% & 44.0\% & 1.44 & 23.9 \\
CheXagent & concise & multi & 38.0\% & 42.0\% & 0.127 & 0.073 & 72.0\% & 12.0\% & 3.96 & 70.9 \\
CheXagent & labels-only & single & 6.0\% & 28.0\% & 0.094 & 0.367 & 12.0\% & 44.0\% & 1.37 & 23.3 \\
CheXagent & labels-only & multi & 32.0\% & 46.0\% & 0.127 & 0.125 & 64.0\% & 28.0\% & 2.92 & 49.7 \\
MedGemma-4B & structured & single & 50.0\% & 68.0\% & 0.239 & 0.047 & 96.0\% & 40.0\% & 0.76 & 30.2 \\
MedGemma-4B & structured & multi & 48.0\% & 64.0\% & 0.188 & 0.053 & 96.0\% & 32.0\% & 0.70 & 56.1 \\
MedGemma-4B & concise & single & 52.0\% & 76.0\% & 0.210 & 0.049 & 100.0\% & 52.0\% & 0.28 & 17.5 \\
MedGemma-4B & concise & multi & 44.0\% & 66.0\% & 0.236 & 0.058 & 88.0\% & 44.0\% & 0.67 & 53.5 \\
MedGemma-4B & labels-only & single & 52.0\% & 70.0\% & 0.252 & 0.047 & 96.0\% & 44.0\% & 0.20 & 7.5 \\
MedGemma-4B & labels-only & multi & 48.0\% & 66.0\% & 0.228 & 0.053 & 92.0\% & 40.0\% & 0.80 & 67.8 \\
MedGemma-27B & structured & single & 50.0\% & 60.0\% & 0.138 & 0.047 & 100.0\% & 20.0\% & 4.86 & 104.3 \\
MedGemma-27B & structured & multi & 46.0\% & 74.0\% & 0.335 & 0.053 & 92.0\% & 56.0\% & 6.27 & 135.1 \\
MedGemma-27B & concise & single & 52.0\% & 66.0\% & 0.263 & 0.051 & 100.0\% & 32.0\% & 3.41 & 72.9 \\
MedGemma-27B & concise & multi & 42.0\% & 62.0\% & 0.211 & 0.085 & 80.0\% & 44.0\% & 4.17 & 87.5 \\
MedGemma-27B & labels-only & single & 50.0\% & 60.0\% & 0.171 & 0.042 & 100.0\% & 20.0\% & 0.45 & 5.8 \\
MedGemma-27B & labels-only & multi & 50.0\% & 68.0\% & 0.248 & 0.056 & 96.0\% & 40.0\% & 3.51 & 73.2 \\
\bottomrule
\end{tabular}
}
\end{table*}

\subsection{Models, Prompts, and Workflows}
We benchmark three medical VLMs: CheXagent-8B \cite{chen2024chexagent}, MedGemma-4B-IT, and MedGemma-27B-IT \cite{medgemma2025}. Including two MedGemma scales lets us separate workflow effects from scaling effects. Each model is tested with three prompt styles:
\begin{itemize}
\item \textbf{Structured}: Detailed, findings-oriented prompts with explicit label constraints. \textit{(e.g., ``Analyze this chest X-ray. Report key findings (lung fields, heart, pleural spaces) and your diagnosis from: [List]. Be concise. Important: Do NOT report a finding unless there is clear evidence for it...'')}
\item \textbf{Concise}: Short prompts with minimal instruction. \textit{(e.g., ``Analyze this chest X-ray. Mention only definite findings and then diagnose using labels from: [List].'')}
\item \textbf{Labels-only}: Restrictive prompts requiring immediate label output with no reasoning. \textit{(e.g., ``Analyze this chest X-ray. Use only labels from: [List]. Return exactly one line and nothing else.'')}
\end{itemize}

Each prompt--model pair is run under two workflows:
\begin{itemize}
\item \textbf{Single-VLM}: Direct, one-shot image-to-label inference.
\item \textbf{Multi-agent}: A four-stage sequential pipeline (illustrated in Figure~\ref{fig:workflow}). (1)~An \emph{extraction} pass asks the VLM to list all visible radiological findings without committing to a diagnosis. (2)~A \emph{radiological formulation} stage maps the extracted findings to candidate diagnostic labels using domain-specific reasoning. (3)~An \emph{internal-medicine formulation} stage re-evaluates the candidates from a clinical perspective, filtering implausible labels. (4)~A \emph{synthesis} stage reconciles the two formulations and outputs the final label set. Each stage re-uses the same frozen model weights; no additional fine-tuning is performed.
\end{itemize}

This design yields 36 experimental configurations: 2 datasets $\times$ 3 models $\times$ 3 prompt styles $\times$ 2 workflows.

\subsection{Evaluation Metrics}
We report six complementary metrics to avoid the single-metric pitfalls discussed in Section~I, alongside \textbf{latency} (s) and \textbf{token count} to quantify computational cost.
\begin{itemize}
    \item \textbf{Exact match}: Fraction of cases where predicted labels exactly equal ground truth.
    \item \textbf{Partial match}: Any non-empty intersection between predicted and ground-truth sets.
    \item \textbf{Macro F1}: Averages per-class F1, penalizing ignorance of rare pathologies.
    \item \textbf{Hallucination FP (Hall FP)}: Mean false-positive rate for absent labels.
    \item \textbf{Normal Accuracy (Normal Acc)}: Accuracy on normal cases (reveals ``Normal'' bias).
    \item \textbf{Abnormal Partial (Abn Partial)}: Partial-match rate isolating pathology sensitivity.
\end{itemize}

\
\
\begin{table*}[t]
\caption{Comprehensive Formal Results on the Private Dataset.}
\label{tab:rih_results}
\centering
\scriptsize
\setlength{\tabcolsep}{4.0pt}
\renewcommand{\arraystretch}{1.08}
\rowcolors{3}{gray!8}{white}
\resizebox{\textwidth}{!}{
\begin{tabular}{lllrrrrrrrr}
\toprule
\textbf{Model} & \textbf{Prompt} & \textbf{Workflow} & \makecell{\textbf{Exact}} & \makecell{\textbf{Partial}} & \textbf{Macro F1} & \makecell{\textbf{Hall FP}} & \makecell{\textbf{Normal Acc}} & \makecell{\textbf{Abn Partial}} & \makecell{\textbf{Latency (s)}} & \textbf{Tokens} \\
\midrule
CheXagent & structured & single & 36.0\% & 78.0\% & 0.436 & 0.147 & 64.0\% & 92.0\% & 0.95 & 13.7 \\
CheXagent & structured & multi & 52.0\% & 62.0\% & 0.317 & 0.038 & 96.0\% & 28.0\% & 4.58 & 54.2 \\
CheXagent & concise & single & 32.0\% & 74.0\% & 0.429 & 0.182 & 56.0\% & 92.0\% & 1.11 & 15.8 \\
CheXagent & concise & multi & 54.0\% & 62.0\% & 0.217 & 0.044 & 100.0\% & 24.0\% & 2.95 & 36.9 \\
CheXagent & labels-only & single & 46.0\% & 82.0\% & 0.457 & 0.129 & 84.0\% & 80.0\% & 1.00 & 11.9 \\
CheXagent & labels-only & multi & 50.0\% & 56.0\% & 0.118 & 0.049 & 100.0\% & 12.0\% & 2.61 & 21.9 \\
MedGemma-4B & structured & single & 52.0\% & 56.0\% & 0.118 & 0.047 & 88.0\% & 24.0\% & 0.35 & 20.0 \\
MedGemma-4B & structured & multi & 50.0\% & 60.0\% & 0.179 & 0.045 & 92.0\% & 28.0\% & 0.63 & 50.4 \\
MedGemma-4B & concise & single & 50.0\% & 56.0\% & 0.120 & 0.051 & 84.0\% & 28.0\% & 0.25 & 13.3 \\
MedGemma-4B & concise & multi & 42.0\% & 54.0\% & 0.158 & 0.058 & 80.0\% & 28.0\% & 0.62 & 48.9 \\
MedGemma-4B & labels-only & single & 52.0\% & 52.0\% & 0.084 & 0.047 & 96.0\% & 8.0\% & 0.18 & 6.4 \\
MedGemma-4B & labels-only & multi & 50.0\% & 56.0\% & 0.128 & 0.049 & 92.0\% & 20.0\% & 0.76 & 64.2 \\
MedGemma-27B & structured & single & 50.0\% & 54.0\% & 0.125 & 0.042 & 100.0\% & 8.0\% & 4.02 & 85.6 \\
MedGemma-27B & structured & multi & 46.0\% & 62.0\% & 0.174 & 0.049 & 88.0\% & 36.0\% & 5.77 & 123.7 \\
MedGemma-27B & concise & single & 50.0\% & 52.0\% & 0.092 & 0.053 & 100.0\% & 4.0\% & 2.88 & 60.8 \\
MedGemma-27B & concise & multi & 36.0\% & 64.0\% & 0.216 & 0.089 & 64.0\% & 64.0\% & 3.46 & 71.8 \\
MedGemma-27B & labels-only & single & 52.0\% & 54.0\% & 0.112 & 0.042 & 100.0\% & 8.0\% & 0.41 & 5.1 \\
MedGemma-27B & labels-only & multi & 52.0\% & 60.0\% & 0.163 & 0.053 & 100.0\% & 20.0\% & 3.25 & 67.1 \\
\bottomrule
\end{tabular}
}
\end{table*}
\begin{figure}[t]
\centering
\includegraphics[width=\columnwidth]{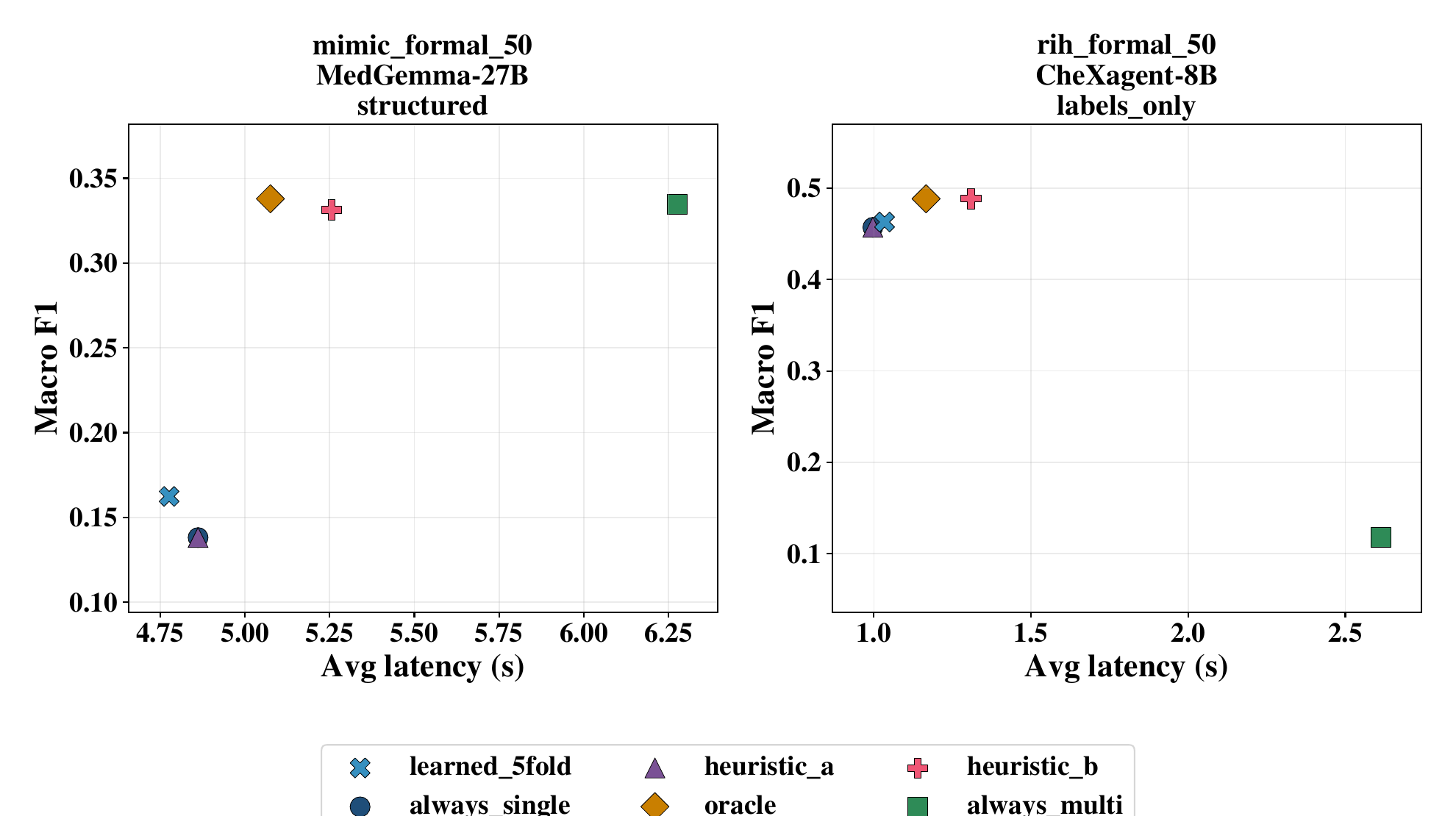}
\caption{Routing trade-off surface: macro F1 versus average latency for different routing strategies.}
\label{fig:routing_tradeoff}
\end{figure}

\begin{figure}[t]
\centering
\includegraphics[width=\columnwidth]{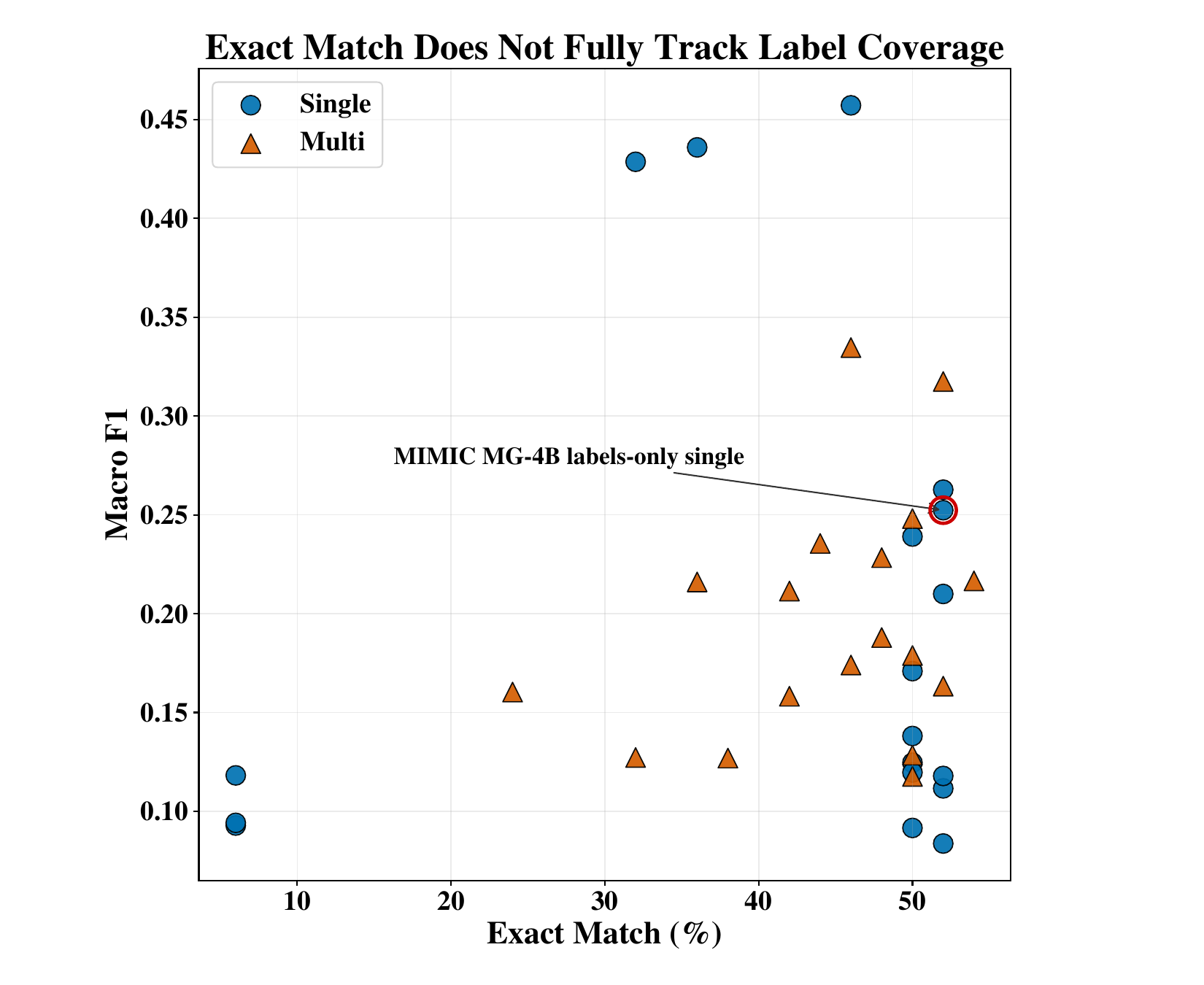}
\caption{Comparison of exact-match and macro F1 scores. High exact-match scores alone can mask poor clinical coverage, because conservative models achieve them by defaulting to ``Normal'' while producing low macro F1.}
\label{fig:exact_vs_f1}
\end{figure}

\begin{figure}[t]
\centering
\includegraphics[width=\columnwidth]{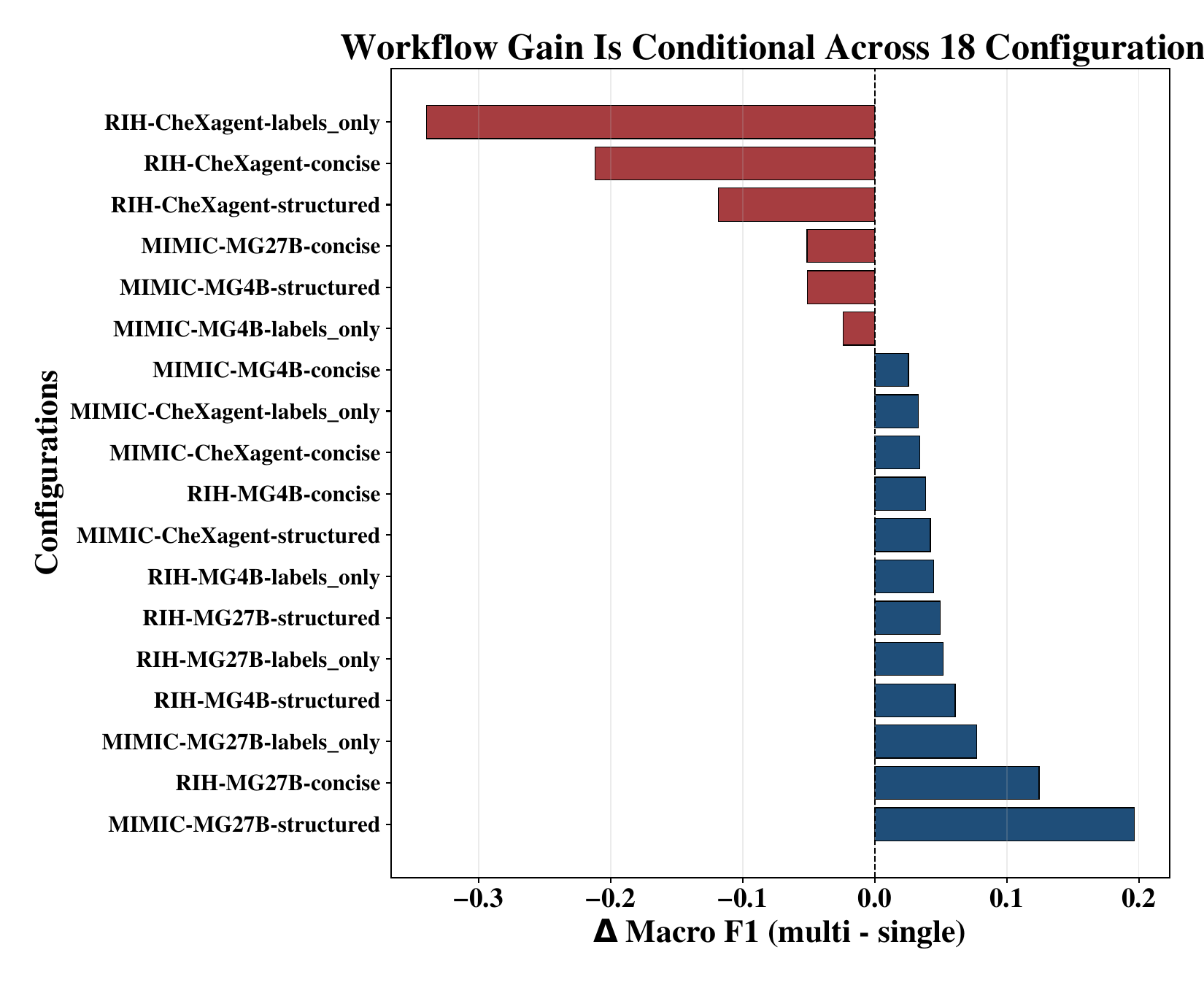}
\caption{Performance difference ($\Delta$ Macro F1) between multi-agent and single-VLM workflows. The direction and magnitude of the change depend strongly on the base model and dataset.}
\label{fig:delta_f1}
\end{figure}

\section{Results and Analysis}

\subsection{Primary Findings}
Our analysis yields three key findings. Detailed performance metrics across all configurations are reported in Table~\ref{tab:mimic_results} (MIMIC) and Table~\ref{tab:rih_results} (Private cohort).

\textbf{1. Model Architecture and Scale Shape Performance.} On the MIMIC dataset, MedGemma-27B with a structured multi-agent pipeline achieved the highest macro F1 (0.335). However, the smaller MedGemma-4B was competitive under direct inference, outperforming MedGemma-27B in single-VLM macro F1 for both structured (0.239 vs.\ 0.138) and labels-only (0.252 vs.\ 0.171) prompts. This shows that performance depends on the interaction between parameter scale and prompt style, not on model size alone.

\textbf{2. Exact Match Does Not Reflect Diagnostic Utility.} MedGemma-4B illustrates this clearly. In the MIMIC labels-only single-VLM setting, it reached 52\% exact match and 70\% partial match with only 0.20s average latency, yet its macro F1 was just 0.252, indicating poor label coverage across pathologies. Accuracy metrics alone fail to expose such gaps (see Figure~\ref{fig:exact_vs_f1}).

\textbf{3. Multi-Agent Reasoning Helps Selectively.} Sequential agentic reasoning substantially improved MedGemma-27B on MIMIC but had little effect on MedGemma-4B and actively hurt CheXagent on the private cohort. The performance difference ($\Delta$ Macro F1) across configurations is shown in Figure~\ref{fig:delta_f1}. Table~\ref{tab:key_claims} summarizes paired bootstrap evidence for these configuration-dependent shifts.

\begin{table*}[t]
\caption{Key comparisons showing configuration-dependent performance shifts and failure types.}
\label{tab:key_claims}
\centering
\small
\setlength{\tabcolsep}{5.2pt}
\renewcommand{\arraystretch}{1.14}
\rowcolors{3}{gray!8}{white}
\resizebox{\textwidth}{!}{
\begin{tabular}{lllrrl}
\toprule
\textbf{Dataset} & \textbf{Comparison} & \textbf{Setting Pair} & \textbf{$\Delta$ Macro F1} & \textbf{$\Delta$ Abn Partial} & \textbf{Primary Failure Signal} \\
\midrule
MIMIC & Multi vs.\ Single & MedGemma-27B, structured & +0.174 & +0.362 & Mitigation of residual normal overcalls \\
Private & Single vs.\ Multi & CheXagent, labels-only & +0.295 & +0.678 & Abnormal-to-normal pathological collapse \\
MIMIC & Model contrast (multi) & MedGemma-27B vs.\ CheXagent, structured & +0.160 & +0.325 & Reduced false-positive propensity in MedGemma-27B \\
\bottomrule
\end{tabular}
}
\end{table*}

\subsection{Typology of Failure Modes}
A closer look at error patterns clarifies why the results vary so much across configurations. We identify two dominant failure mechanisms.

\textbf{Abnormal-to-normal collapse.} A model labels a genuinely abnormal case as ``Normal,'' missing all pathology. On the MIMIC subset, MedGemma-27B (structured, single-VLM) collapsed abnormal cases in this way 30\% of the time. The multi-agent pipeline reduced this rate to 10\%, because the additional extraction and formulation stages recovered findings that the single pass missed. MedGemma-4B, however, did not benefit similarly: it already achieved low collapse rates under single-VLM inference, so the multi-agent overhead introduced noise without compensating gains.

\textbf{Over-regularization by multi-agent consensus.} When the pipeline stages disagree, the synthesis step tends to default to ``Normal'' as a conservative fallback. This effect was most pronounced for CheXagent on the private dataset (labels-only): the single-VLM collapse rate of 10\% rose to 44\% after adding the multi-agent pipeline. Because CheXagent generates shorter, less structured outputs, the extraction stage often yields sparse evidence, leading downstream stages to discount legitimate but weakly expressed findings.

These two mechanisms act in opposite directions. For large models with verbose outputs (MedGemma-27B), multi-agent processing rescues missed findings. For smaller or more terse models (MedGemma-4B, CheXagent), the same pipeline strips away genuine detections. This asymmetry is the root cause of the configuration-dependent routing gains reported in Section~III-E.

\subsection{Qualitative Case Examples}
Two representative cases illustrate the dual-edged nature of multi-agent pipelines.

\textbf{Case 1: Improvement (MIMIC).} In \texttt{mimic\_formal\_003}, the ground truth is \texttt{Cardiomegaly, Edema, Pleural Effusion}. MedGemma-27B (single-VLM) misses all findings and outputs \texttt{Normal}. The multi-agent workflow correctly identifies \texttt{Cardiomegaly, Edema}, turning a complete miss into a useful partial match.

\textbf{Case 2: Regression (Private).} In \texttt{private\_formal\_043}, the report notes stable cardiomegaly with clear lungs. CheXagent (labels-only, single-VLM) correctly predicts \texttt{Cardiomegaly}. The multi-agent workflow collapses the same case to \texttt{Normal}, illustrating the abnormal-to-normal suppression pattern seen in the aggregate statistics.

\subsection{Decision-Time Routing Proof-of-Concept}

\begin{table*}[t]
\caption{Evaluation of Decision-Time Routing Topologies Across Core Representative Settings.}
\label{tab:routing_poc}
\centering
\scriptsize
\setlength{\tabcolsep}{4.5pt}
\renewcommand{\arraystretch}{1.08}
\rowcolors{3}{gray!8}{white}
\resizebox{\textwidth}{!}{
\begin{tabular}{llrrrr}
\toprule
\textbf{Configuration} & \textbf{Routing Strategy} & \textbf{Macro F1} & \textbf{Latency (s)} & \textbf{Tokens} & \textbf{Multi-Agent Esc. (\%)} \\
\midrule
MIMIC / MedGemma-27B / structured & always-single & 0.138 & 4.86 & 104.3 & 0.0 \\
MIMIC / MedGemma-27B / structured & always-multi & 0.335 & 6.27 & 135.1 & 100.0 \\
MIMIC / MedGemma-27B / structured & \textbf{oracle} & \textbf{0.338} & 5.08 & 109.0 & 20.0 \\
MIMIC / MedGemma-27B / structured & heuristic A & 0.138 & 4.86 & 104.3 & 0.0 \\
MIMIC / MedGemma-27B / structured & heuristic B & 0.331 & 5.26 & 112.5 & 40.0 \\
MIMIC / MedGemma-27B / structured & learned (5-fold LR) & 0.162 & 4.78 & 102.2 & 10.0 \\
\midrule
Private / CheXagent / labels-only & always-single & 0.457 & 1.00 & 11.9 & 0.0 \\
Private / CheXagent / labels-only & always-multi & 0.118 & 2.61 & 21.9 & 100.0 \\
Private / CheXagent / labels-only & \textbf{oracle} & \textbf{0.488} & 1.17 & 13.6 & 10.0 \\
Private / CheXagent / labels-only & heuristic A & 0.457 & 1.00 & 11.9 & 0.0 \\
Private / CheXagent / labels-only & \textbf{heuristic B} & \textbf{0.488} & 1.31 & 15.0 & 18.0 \\
Private / CheXagent / labels-only & learned (5-fold LR) & 0.463 & 1.03 & 12.0 & 2.0 \\
\bottomrule
\end{tabular}
}
\end{table*}

The configuration-dependent gains described above motivate an inference routing approach (Figure~\ref{fig:routing_tradeoff}): use features available from the initial single-VLM pass to decide whether multi-agent escalation is worthwhile. We tested four routing strategies on two representative configurations (MIMIC/MedGemma-27B/structured and Private/CheXagent/labels-only): static always-single, static always-multi, a hindsight oracle, two lightweight heuristics, and a supervised logistic regression router. \textbf{Heuristic A} triggers multi-agent escalation only when the single-VLM predicts ``Normal'' on a case flagged as complex (severity level 2). \textbf{Heuristic B} escalates whenever the single-VLM output has zero partial-match overlap with the ground-truth labels.

The learned router uses features from the single-VLM pass (predicted label count, normality flag, token count, severity level, and partial-match indicators). The training label is \texttt{use\_multi}=1 when the multi-agent macro F1 exceeds the single-VLM macro F1 for that case. We train a cross-validated logistic regression model (\texttt{max\_iter=500}, 5-fold) and aggregate predictions.

Table~\ref{tab:routing_poc} shows that routing consistently outperforms fixed workflows in balancing cost and diagnostic quality. On MIMIC, always-multi is a strong baseline (macro F1 0.335), but the oracle matches it (0.338) while escalating only 20\% of cases. On the private dataset, where always-single (0.457) far exceeds always-multi (0.118), simple heuristics and the oracle both reach 0.488 with selective escalation.

Two observations deserve further comment. First, Heuristic~A escalates zero cases on MIMIC because MedGemma-27B already tends to predict ``Normal'' for most ambiguous inputs. Since the heuristic conditions on predicting ``Normal'' \emph{and} high pre-assessed complexity, the conjunction is rarely satisfied, making the rule too conservative for this model--dataset pair. Heuristic~B, which uses a purely output-based trigger (empty partial-match overlap), avoids this issue and achieves near-oracle performance. Second, the learned router is limited by the small training set (50 cases per fold). With only 5--10 positive routing labels per fold, logistic regression cannot reliably separate the two classes; we expect this gap to narrow with larger cohorts.

\section{Discussion and Future Directions}
Our results show that chest X-ray VLM reliability depends on the joint interaction of model family, parameter scale, and workflow design. Smaller models perform well under direct inference but gain little---or even lose ground---from multi-agent deliberation that benefits larger counterparts. Hallucination rates reinforce this pattern: CheXagent in single-VLM mode produced Hall FP rates above 0.35 on MIMIC (Table~\ref{tab:mimic_results}), which the multi-agent pipeline reduced below 0.13, whereas MedGemma models already had low hallucination rates ($<$0.06) and sometimes lost true positives from the extra filtering. No single, fixed inference strategy is therefore optimal across configurations.

Decision-time routing is a first step toward adaptive deployment. Future work should extend this framework to federated, continuously learning routers that adapt to institutional disease prevalence and demographic variation, and to multi-modal inputs (e.g., EHR timelines combined with imaging) where routing decisions become richer.

\textbf{Limitations.} Both cohorts contain only 50 cases, limiting statistical power and the learned router's training signal. The label space omits rare diagnoses (e.g., pneumomediastinum) that may expose additional failure modes. All experiments use frozen weights and greedy decoding; sampling strategies or prompt-tuning could alter the configuration rankings.

\section{Conclusion}
We evaluated chest X-ray VLMs across 36 workflow configurations and found that blanket multi-agent deployment is not always best. Multi-agent reasoning substantially improved MedGemma-27B on public data, but simpler direct-prompting setups with CheXagent outperformed it on our private cohort, and the compact MedGemma-4B remained competitive under single-VLM inference. These results show that aggressive prompt chaining can over-regularize predictions and suppress real pathology. Our decision-time routing heuristic addresses this by escalating to multi-agent inference only when needed, approaching oracle-level performance while keeping latency low. For safe and scalable clinical deployment, VLM systems should move beyond fixed workflows toward dynamic, cost-aware routing, systematically reducing unnecessary token costs while maintaining diagnostic integrity.

\bibliographystyle{IEEEtran}
\bibliography{chase2026_workshop_refs}

\end{document}